\documentclass[aps,twocolumn]{revtex4}

\usepackage{amssymb}
\usepackage{amsmath}
\usepackage{graphicx}
\newcommand{\be}{\begin{equation}}
\newcommand{\ee}{\end{equation}}
\newcommand{\ben}{\begin{eqnarray}}
\newcommand{\een}{\end{eqnarray}}
\newcommand{\lb}{\label}

\begin{document}

\title{Finite-size effects on the phase diagram of difermion condensates in
two-dimensional four-fermion interaction models}

\author{L. M. Abreu$^{1}$, A. P. C. Malbouisson$^{2}$,
J. M. C. Malbouisson$^{1}$ }

\affiliation{$^{1}$Instituto de F{\'i}sica, Universidade Federal da
Bahia, 40210-340, Salvador, BA, Brazil}

\affiliation{$^{2}$Centro Brasileiro de Pesquisas F{\'i}sicas, MCT,
22290-180, Rio de Janeiro, RJ, Brazil}

\begin{abstract}

We investigate finite-size effects on the phase structure of
chiral and difermion condensates at finite temperature and density in the
framework of the two-dimensional large-$N$ Nambu-Jona-Lasinio model. We  take into account size-dependent effects by making use of
$zeta$-function and compactification methods. The thermodynamic
potential and the gap equations for the chiral and difermion condensed phases are
then derived in the mean-field approximation. Size-dependent
critical lines separating the different phases are obtained considering anti-periodic boundary conditions for the spatial coordinate.

\end{abstract}

\maketitle

\section{Introduction}

Investigation on the phase structure of the strongly interacting
matter is one of the most interesting topics in the realm of the
standard model for the fundamental forces, in particular for the
confining-deconfining phase transition. This transition is expected
to take place far from the asymptotic freedom (high energy) domain
of Quantum Chromodynamics (QCD) and so, non-perturbative methods are
needed in order to approach it and other phenomena occurring in this
region. The most currently used non-perturbative methods involve
computer simulations in lattice field theory and have given many
interesting results, as for instance numerical estimates for the
confining-deconfining transition temperature. In what concerns
precise analytical studies, they are very difficult in the low
energy domain of QCD, due to its complex field-theoretical
structure.

In reason of these difficulties, phenomenological models for QCD
have been adopted along the years. Simplified effective theories
have been largely employed as a laboratory to get, analytically,
insights on the behavior of hadronic matter, particularly in their
low-dimensional versions. One of the most frequently used of these
models is the Nambu-Jona-Lasinio (NJL) model~\cite{NJL}. It is a
theory of nucleons and mesons, defined in a spacetime with an even
number of dimensions, constructed from directly interacting Dirac
fermions with chiral symmetry. Phase transition is considered  in a
manner analogous to the appearance of Cooper pairs from electrons in
the BCS theory of superconductivity. This is the origin of the
expression "color superconductivity" for the hadronic phase
transition in the context of this model. Nowadays it is well known
the usefulness of the NJL model in the description of the phase
diagram of both chiral broken phase (quark-antiquark condensation)
and color superconducting phase (diquark condensation). In addition,
the NJL model is specially convenient in the study of systems under
certain conditions, like finite temperature~\cite{CMC,CCMMS}, finite chemical
potential, external gauge field, among others~\cite{Kl,HK,Bu}.

An interesting aspect in the study of the phase transitions of the
NJL model is the relevance of the fluctuations due to finite-size
effects in the phase diagram. With this purpose, different
approaches have been used to study various aspects of these
effects~\cite{SK,DS,Kim,HCGL,VVK,KS,BS,GG,KKK,EKTZ,KH,AGS,AMMS}.
Other phenomenological approaches have also been adopted. For
instance in~\cite{wilczek} a variational procedure is employed to
study finite density QCD in a  model in which the interaction
between quarks is supposed to be mediated by instantons. This is
related to the picture of hadrons as assumed in the MIT bag model,
where nucleons are considered as droplets in a chirally symmetric
restored phase. These authors find that at densities high enough the
chirally symmetric phase fills space, and color symmetry is broken
by the formation of a quark-quark condensate.

In this paper we intend to generalize,
by including finite size effects, previous results obtained for
finite temperature~\cite{CMC,CCMMS}. We make use of the techniques introduced in
Refs.~\cite{AGS,AMMS} and investigate finite-size effects on the
phase structure of chiral and difermion condensates at finite temperature and density in the framework of the two-dimensional large-$N$ NJL model.
 This is done through $zeta$-function regularization and compactification
methods~\cite{EE,Ademir}. This approach allows to determine analytically
the size-dependence of the effective potential and the gap equation.
Then, phase diagrams at finite temperature and chemical potential,
where the symmetric and nonsymmetric phases are separated by
size-dependent critical lines, are obtained.

Let us remark that our interest in the two-dimensional version of
the NJL model is an attempt to investigate the qualitative aspects
relevant to the chiral and difermion condensations under the influence of size finiteness of the system. In fact many properties of four-fermion
interacting models are similar in lower and higher dimensions, and
so, we can expect that results obtained in the 2D NJL model reflect
properties of a more realistic 4D model. 

It could be argued that
spontaneous symmetry breaking does not occur in two dimensions, as a
consequence of the Mermin-Wagner-Coleman theorem. However this is
not the case in the present situation. We emphasize that although
the Mermin-Wagner-Coleman theorem denies the spontaneous breaking of
continuous symmetries in two dimensions~\cite{MWC}, it does not
apply in the large-$N$ limit~\cite{SK,DS,EKTZ,BCMP,CMC,CCMMS,ST,KO,MBC,KPR,BDT}.
Therefore since our interest is in the analysis of the possible
breaking of continuous symmetries, it is legitimate to
study symmetry breaking effects in terms of the low-dimensional NJL
model in the large-$N$ limit.

We organize the paper as follows. In Section~II, we calculate the
effective potential of the NJL model in the mean-field
approximation, using the $zeta$-function method. The analysis of
spontaneous symmetry breaking induced by the chiral and difermion condensates are
also done. The size-dependent gap equations is discussed in
Section~III, while the phase diagrams are shown and analyzed in
Section~IV. Finally, Section~V presents some concluding remarks.

\section{The model}

Our starting point is the two-dimensional massless version of the
extended NJL model, described by the Lagrangian density~\cite{CMC,CCMMS,MBC},
\ben
{\cal L} & = &  \bar{\psi}^{(i)} \;\left(  i \not{\!\partial} - \mu \gamma ^0 \right) \; \psi ^{(i)} +
\frac{g_S}{2} \left(
\bar{\psi}^{(i)}  \psi ^{(i)} \right)^2 \nonumber \\
& & + g_D \left( \bar{\psi}^{(i)} \gamma _5 \psi ^{(j)}
\right)\left( \bar{\psi}^{(i)} \gamma _5 \psi ^{(j)} \right),
\label{L1}
\een
where $\psi $ and $\bar{\psi}$ are the fermion fields carrying $N$ flavors ($i,j = 1,...,N$; repeated flavor indices are summed), $\mu$ is the chemical potential and
the $\gamma$ matrices are in the two-dimensional
space representation, with $\gamma ^5 = \gamma ^0 \gamma ^1$. Notice that the Lagrangian density possess $O(N)$ flavor symmetry and discrete chiral symmetry. In the following, unless explicitly stated, we use natural units, $\hbar\,=\,k_B\,=\,c\,=1$.

Choosing a particular representation we have,
\be \gamma ^0 = \left(
\begin{array}{cc}
  0 & 1 \\
  1 & 0 \\
\end{array}
            \right),
\gamma ^1 = \left(
              \begin{array}{cc}
                0 & -1 \\
                1 & 0 \\
              \end{array}
            \right), 
\label{gamma}
\ee
where, in this representation, $C=-\gamma ^1$. In this case, the pairing term reads
\ben
g_D \left( \bar{\psi}^{(i)} \gamma _5 \psi ^{(j)} \right)\left(
\bar{\psi}^{(i)} \gamma _5 \psi ^{(j)} \right) = \nonumber \\
- \frac{g_D}{2} \left( \varepsilon _{\alpha \beta} \psi _{\alpha
}^{\dagger (i)} \psi _{\beta }^{\dagger (i)} \right) \left(
\varepsilon _{\gamma \delta} \psi _{\gamma }^{ (j)}  \psi _{\delta
}^{ (j)} \right) . \label{pairing}
\label{pairing}
\een

We perform the bosonization by introducing the auxiliary fields $\sigma, \Delta $, associated to the bilinear forms in the above Lagrangian density as:  $g_s \bar{\psi}^{(i)}  \psi ^{(i)} \equiv \sigma$ and
$g_d \varepsilon _{\gamma \delta} \psi _{\gamma }^{ (i)}  \psi _{\delta }^{ (i)} \equiv
\Delta $. Therefore the modified Lagrangian density becomes,
\ben
 \widetilde{{\cal L}} & = &   \bar{\psi}^{(i)} \;\left(  i \not{\!\partial} -
\sigma - \mu \gamma^0 \right)\;  \psi ^{(i)} - \frac{1}{2} \Delta ^{\dagger} \left( \varepsilon _{\gamma \delta} \psi _{\gamma }^{ (i)}  \psi _{\delta }^{ (i)}\right) \nonumber \\
& & + \frac{1}{2} \left( \varepsilon _{\alpha \beta} \psi _{\alpha }^{\dagger (i)}  \psi _{\alpha }^{\dagger (i)}
\right) \Delta -
\frac{1}{2 g_S} \sigma ^2 - \frac{1}{2 g_D} |\Delta | ^2.
\label{L2}
\een
We see from Eq.~(\ref{L2}) that the auxiliary field
$\sigma$ plays the role of a dynamical fermion mass, such that when
it has a non-vanishing value, the system is in the chiral broken
phase. The auxiliary field $\Delta$ is associated with the difermion condensate.

Then, integration over $ \psi $ and $ \psi ^{\dagger }$ generates the effective action,
\ben
\Gamma _{eff} (\sigma, \Delta) & = & \int d²x \left( -\frac{1}{2 g_S} \sigma ^2 - \frac{1}{2 g_D} |\Delta | ^2 \right) \nonumber \\
& & - \frac{i}{2} \rm{Tr} \ln{D} ,
\label{eff_action}
\een
where
\ben
D = \left(
\begin{array}{cc}
  -h & \gamma ^1 \Delta ^{\dagger}\\
   -\gamma ^1 \Delta & h^{T} \\
\end{array}
            \right),
\label{D_op}
\een
with
\ben
h & = & i \partial _0 + i \gamma ^5 \partial _1 - \mu - \sigma \gamma ^0, \nonumber \\
h ^T & = & -i \partial _0 - i \gamma ^5 \partial _1 - \mu - \sigma \gamma ^0.
\label{op_h}
\een

Notice that the trace over the flavor indices in Eq. (\ref{eff_action}) gives a factor $N$, which allow us to set, in the large-$N$ limit, $g_S N = G_S$ and $g_D N = G_D$, with $G_S$ and $G_D$ fixed at $N\rightarrow \infty$. Thus, the
effective potential is obtained in the mean-field approximation (i. e., $\sigma$ and $|\Delta|$ uniform) from Eq. (\ref{eff_action}),
\ben
U_{eff} (\sigma, \Delta)  & = &   \frac{ \sigma ^2}{2 G_S} +
\frac{|\Delta | ^2 }{2 G_D} + \frac{i}{2}  {\rm tr} \ln{(h^T h)}\nonumber \\
& & + \frac{i}{2}  {\rm tr} \ln{\left[ 1 - |\Delta|^2 (h ^T)^{-1} \gamma ^1 (h)^{-1}\gamma ^1\right]},
\label{pot1}
\een
where $\rm{tr}$ means the trace over spinor indices.

Our aim is to take into account simultaneously finite-temperature and finite-size effects on
the phase structure of the model; in order to do this, we consider an Euclidean space, with imaginary time and the spatial
coordinate being compactified. We denote the Euclidian coordinate
vectors by  $x_E = (x_0, x)$, with $x_0 \in [0, \beta] $ and $ x
\in [0, L]$, where $\beta$ is the inverse temperature, $\beta = T^{-1}$, and
$L$ is the size of the system. This corresponds to the generalized
Matsubara prescription,
\ben
\int \frac{d^2 k }{(2 \pi)^2} f(k_0,k) & \rightarrow & \frac{1}{\beta L}
\sum_{n_0,n = -\infty}^{\infty} f(\omega_{n_0},\omega_{n}), \nonumber
\een
where
\ben
k_0 & \rightarrow & \omega_{n_0} = \frac{2 \pi }{\beta} \left( n_0 + \frac{1}{2} \right);\;
n_0 = 0,\pm 1, \pm 2,\cdots, \nonumber \\
k & \rightarrow & \omega_{n} = \frac{2 \pi }{L} \left( n + c \right);\;\;\; n =
0,\pm 1, \pm 2, ...;
\nonumber
\label{Matsubara}
\een
in the above equation the quantity $c$ is such that, $c = 0$ and $c = \frac{1}{2} $ for,    respectively, periodic and  antiperiodic spatial boundary conditions. In this article we will restrict ourselves to antiperiodic spatial boundary conditions, which is a natural choice for fermionic systems. The case of periodic spatial boundary conditions would follow along parallel lines. Unless explicitly stated, in all cases studied the spatial boundary conditions are antiperiodic.

Using Eq.~(\ref{Matsubara}) we get, after some manipulations, the effective potential
carrying finite-temperature and finite-size effects,
omitting terms independent of $|\Delta|$ and  $\sigma$,
\ben
U_{eff}^{\beta,L} (\sigma, \Delta)& = &  \frac{\sigma ^2 }{2 G_S} + \frac{|\Delta| ^2 }{2 G_D}
\nonumber \\
& & -  \frac{1}{2 \beta L} \sum _{\pm}  \sum_{n_0, n = -\infty}^{\infty}
\ln{\left[ \omega_{n_0} ^2 + k_{\pm} ^2 \right]},
\label{pot2sigma}
\een
where
\ben
k_{\pm} ^2 & = & |\Delta| ^2 + \sigma ^2 + \mu ^2 + \omega _{n} ^2 \nonumber \\
& & \pm 2 \left[ |\Delta| ^2 \sigma ^2 + \mu ^2 \left( \omega _{n} ^2 +  \sigma ^2 \right) \right] ^{\frac{1}{2}}.
\label{var}
\een

The effective potential in Eq.~(\ref{pot2sigma}) can be rewritten in
terms of the Epstein $zeta$-functions, $Y(s)$, that is,
\ben
U_{eff} ^{\beta,L} (\sigma, \Delta) & = &\frac{\sigma ^2 }{2 G_S} + \frac{|\Delta| ^2 }{2 G_D} + \frac{1}{2 \beta L}
\sum _{\pm}  \frac{d}{d s }\left. Y _{\sigma,\Delta} ^{\pm} (s) \right| _{s=0}, \nonumber\\
\label{pot3}
\een
where
\ben
Y _{\sigma,\Delta}  ^{ \pm} (s) & =& \sum_{n_0,n = -\infty}^{\infty}
\left[\omega_{n_0} ^2 + k_{\pm} ^2 \right]^{-s } .
\label{epstein1}
\een

The analysis of the phase diagram of the temperature and boundary-dependent model is performed
through the solutions of the gap equation containing thermal and boundary corrections. However, for completeness
and to set up the free space parameters, in the next section we start by treating the zero-temperature model without chemical potential and in absence of spatial boundaries.

\section{Model at $T=0,\, \mu = 0$ and without spatial boundaries}
\label{model-free}

Let us study the model introduced in the previous section, without compactification of the spatial dimension and at zero temperature. This case has been well-studied in Ref. \cite{CCMMS}; here we perform this study to define the zero-temperature free space parameters in absence of chemical potential.
The renormalization conditions for the coupling constants are,
\ben
\frac{1}{ G_{SR}}  & =&  \left.\frac{\partial ^2  }{\partial
\sigma ^2}U_{eff}  (\sigma, \Delta) \right|_{ \sigma = \sigma _0, \Delta = \Delta _0
}  ,
\lb{rencond1}
\een
and
\ben
\frac{1}{ G_{DR}}  & =&  \left.\frac{\partial ^2  }{\partial
\Delta ^2}U_{eff}  (\sigma, \Delta) \right|_{ \sigma = \sigma _0, \Delta = \Delta _0  }, 
\lb{rencond}
\een
where $G_{SR}$ and $G_{DR}$ are the renormalized coupling constants, and $\sigma_0$ and $\Delta _0$ are scale parameters. Then, taking $\mu\,= T\,=0$ in  Eq.~(\ref{pot1}), and performing the integration over $p_0$, the unrenormalized effective potential can be rewriten as
\ben
U_{eff} (\sigma, \Delta)  & = &   \frac{ \sigma ^2}{2 G_S} +
\frac{|\Delta | ^2 }{2 G_D} - \frac{1}{2 \pi} \int dk \left( k_{+} + k_{-} \right) ,\nonumber \\
\label{unren-effec}
\een
where $k_{\pm}$ is given by Eq. (\ref{var}), with the replacement $\omega _n \rightarrow k$. So, we obtain from (\ref{rencond1}) and (\ref{rencond}) the following relations \cite{CMC,CCMMS,MBC}
\ben
\frac{1}{ G_{SR}}  & =&  \frac{1}{ G_{S}} + \frac{1}{ \pi} - X,\nonumber \\
\frac{1}{ G_{DR}}  & =&  \frac{1}{ G_{D}} + \frac{1}{4 \pi} - \frac{1}{8 \pi} \frac{\sigma_0}{\Delta _0} \ln{\left| \frac{\Delta _0 - \sigma _0 }{\Delta _0 + \sigma _0 }\right| } - \frac{X}{2},
\lb{ren_coupl}
\een
where the quantity $X$ carries the divergent part,
\ben
X & = & \frac{1}{2 \pi} \int dk \left\{ [k ^2 + (\sigma _0 + \Delta _0)^2]^{- \frac{1}{2}} \nonumber\right.  \\
&  & \left. +  [k ^2 + (\sigma _0 - \Delta _0)^2]^{- \frac{1}{2}} \right\} \nonumber \\
& = & \frac{1}{\pi}  \left\{ \frac{1}{\varepsilon} + \ln{\left[\frac{2}{\sqrt{|\sigma _0 ^2 - \Delta _0 ^2|}} \right] }  \right\},
\label{divergence}
\een
for $\varepsilon \rightarrow 0$. 

Hence now we are able to write the renormalized effective potential,
\ben
\bar{U}_{eff} (\sigma, \Delta)  & = &   \alpha _1 \sigma ^2 +
\alpha _2 |\Delta | ^2 \nonumber \\
& & -   \frac{1}{2 } {\rm FP}\left\{ \sum _{\pm} \int \frac{d^{2}k}{(2 \pi)^2}
\ln{\left[ k_{n_0} ^2 + k_{\pm} ^2 \right]} \right\}, \nonumber \\
\label{ren-effec}
\een
where
\ben
\alpha_1 & =& \frac{1}{2 G_{SR}} - \frac{1}{2 \pi} , \nonumber \\
\alpha_2 & =& \frac{1}{2 G_{DR}} - \frac{1}{4 \pi} + \frac{1}{8 \pi} \frac{\sigma_0}{\Delta _0} \ln{\left| \frac{\Delta _0 - \sigma _0 }{\Delta _0 + \sigma _0 }\right| } ;
\label{alpha}
\een
in the above equation, ${\rm FP}\left\{ ... \right\}$ means the
finite part of the terms between brackets.  When we set $\alpha _1 = 0$, we have the vacuum values, $\sigma \neq 0 $ and $\Delta =0$. Choosing renormalization scales as $\sigma _0 = m $ ($m$ is a scale parameter) and $\Delta _0 = 0$, then $\alpha_2 > 0 $, with $G_{SR} = \pi$. This is the chiral condensate sector. On the other hand, if we take the vacuum with $\Delta \neq 0 $ and $ \sigma=0$, and choosing $\alpha_2 = 0 $, $\alpha_1 > 0 $, $\Delta _0 = m $ and $\sigma _0 = 0$, we have $G_{DR} = 2 \pi$, corresponding to the difermion condensate sector.

\section{Model at finite $T, \mu$ and with spatial boundaries}

Now we take into account temperature, chemical potential and finite-size dependence. Noticing that the $T,\mu, L$-dependent contributions do not alter the structure of ultraviolet divergences discussed in Section \ref{model-free}, then the renormalized effective potential is given by
\ben
\bar{U}_{eff} ^{\beta,L} (\sigma, \Delta) & = &\alpha _1 \sigma ^2 +
\alpha _2 |\Delta | ^2 \nonumber \\
& &  + \frac{1}{2 \beta L}
\sum _{\pm}{\rm FP}\left\{  \frac{d}{d s }\left. Y _{\sigma,\Delta} ^{\pm} (s) \right| _{s=0}  \right. \nonumber \\
& & \left.- Y _{\sigma,\Delta} ^{\pm} (s) \ln{m^2}\right\}.
\label{pot3}
\een
Thus, the $T,\mu, L$-dependent phase diagram of the model can be analyzed from $\bar{U}_{eff} ^{\beta,L}$ and the gap equations.

\subsection{The chiral condensate sector}

To analyze the pure chiral condensate sector of the model, we must take the effective potential in Eq.~(\ref{pot3}) with $\Delta = 0$ and $\alpha_1 = 0$, as pointed out at the end of Section \ref{model-free}. In order to do this, we perform an analytical continuation of the Epstein $zeta$-function $Y_{\sigma} ^{ \pm}$ \cite{AGS,AMMS,EE}, which gives, 
\ben
\bar{U}_{eff} ^{\beta,L} (\sigma) & = & \frac{1}{4 \pi} \sigma ^2 [\ln{\sigma ^2} -1 ]
\nonumber \\
& &  + \frac{2 \sigma}{\pi L} \sum _{n = 1}^{\infty} \frac{\cos{(2 \pi n c)}}{n} K_{1}(n L \sigma) \nonumber \\
& & -\frac{2}{\beta L} \sum _{\pm} \sum _{n = - \infty}^{\infty} \ln{\left\{ 1 + e^{- \beta \left[ (\omega _n ^2 + \sigma ^2)^{\frac{1}{2}} \pm \mu \right]} \right\}}. \nonumber \\
\label{pot4}
\een
Remember that we have fixed $m$ as
the scale of the model, redefining the relevant quantities as
$\bar{U}_{eff} / m^2 \rightarrow \bar{U}_{eff}$, $\sigma / m  \rightarrow \sigma $, $L m
\rightarrow L$, $ \mu / m  \rightarrow \mu$ and $\beta m
\rightarrow \beta$.
Firthermore, we can see that in the bulk limit, $L \rightarrow \infty$, Eq.~(\ref{pot4}) becomes
\ben
\bar{U}_{eff} ^{\beta,L} (\sigma) & = & \frac{1}{4 \pi} \sigma ^2 [\ln{\sigma ^2} -1 ]
\nonumber \\
& & -\frac{2}{\beta } \sum _{\pm} \int \frac{dk }{2 \pi} \ln{\left\{ 1 + e^{- \beta \left[ ( k ^2 + \sigma ^2)^{\frac{1}{2}} \pm \mu \right]} \right\}},  \nonumber \\
\label{pot5}
\een
an expression which agrees to that in Ref. \cite{CCMMS}.

The ground state is analyzed by means of the minimum of the effective potential, which corresponds to the gap equation, 
\ben
\left. \frac{\partial}{\partial \sigma}  \bar{U}_{eff} ^{\beta,L} (\sigma) \right. = 0.
\label{gap0s}
\een
The non-vanishing solution of the gap equation, yields the dynamically generated fermion mass, which comes from the equation, 
\ben
& & \frac{1}{ \pi} \ln{\sigma ^2} - \frac{2}{\pi L} \sum _{n = 1}^{\infty} \cos{(2 \pi n c)} K_{0}(n L \sigma) \nonumber \\
& &  + \frac{2}{L} \sum _{\pm} \sum _{n = - \infty}^{\infty}
\frac{1}{ (\omega _n ^2 + \sigma ^2)^{\frac{1}{2}}} \frac{1}{  e^{ \beta \left[ (\omega _n ^2 + \sigma ^2)^{\frac{1}{2}} \pm \mu \right]} +1 }  =0 .  \nonumber \\
\label{gap1s}
\een

\subsection{The difermion condensate sector}

On the other hand, the pure chiral condensate sector is studied by taking in Eq.~(\ref{pot3}) $\sigma = 0$ and $\alpha_2 = 0 $. After performing the analytical continuation of the Epstein $zeta$-function, $Y_{\sigma} ^{ \pm}$, we have the following effective potential,
\ben
\bar{U}_{eff} ^{\beta,L} (\Delta) & = & \frac{1}{4 \pi} \Delta ^2 [\ln{\Delta ^2} -1 ]
\nonumber \\
& &  + \frac{2 \Delta}{\pi L} \sum _{n = 1}^{\infty} \frac{\cos{(2 \pi n c)}}{n}
\cos{( \mu L n )} K_{1}(n L \Delta) \nonumber \\
& & -\frac{2}{\beta L} \sum _{\pm} \sum _{n = - \infty}^{\infty} \ln{\left\{ 1 + e^{- \beta \left[ \left( \omega _n \pm  \mu \right) ^2 + \Delta ^2 \right]^{\frac{1}{2}} } \right\}},  \nonumber \\
\label{pot6}
\een
where  we have used again $m $ as 
the scale parameter, redefining
$\bar{U}_{eff} / m^2 \rightarrow \bar{U}_{eff}$, $\Delta / m  \rightarrow \Delta $, $L m  \rightarrow L$, $ \mu / m  \rightarrow \mu$ and $\beta m
\rightarrow \beta$.

To verify the consistency of the model, we take the situation without spatial boundaries, that is, $L \rightarrow \infty$. In this case, Eq. (\ref{pot6}) becomes
\ben
\bar{U}_{eff} ^{\beta,L} (\Delta) & = & \frac{1}{4 \pi} \Delta ^2 [\ln{\Delta ^2} -1 ]
\nonumber \\
& & -\frac{2}{\beta } \sum _{\pm} \int \frac{dk }{2 \pi} \ln{\left\{ 1 + e^{- \beta \left[  k ^2 + \sigma ^2\right]^{\frac{1}{2}} } \right\}},
\label{pot7}
\een
which is independent of the chemical potential, in agreement with Ref. \cite{CCMMS}.

The gap equation for the difermion sector is given by
\ben
\left. \frac{\partial}{\partial \Delta}  \bar{U}_{eff} ^{\beta,L} (\Delta) \right. = 0,
\label{gap0d}
\een
from which we get the $(\beta, \mu , L)$-dependent non-vanishing solution in the form, 
\ben
& & \frac{1}{ \pi} \ln{\Delta ^2} - \frac{2 }{\pi L} \sum _{n = 1}^{\infty} \cos{(2 \pi n c)} \cos{( \mu L n )} K_{0}(n L \Delta) \nonumber \\
& & -  \frac{2}{L} \sum _{\pm} \sum _{n = - \infty}^{\infty} \sum _{n = 1 }^{\infty} (-1)^{n} \left( \frac{n \beta}{ 2 \pi\sqrt{ \left( \omega _n \pm  \mu \right)^2 + \Delta ^2} } \right)^{\frac{1}{2}}  \nonumber \\
& & \times K _{\frac{1}{2}} \left( n \beta  \sqrt{ \left( \omega _n \pm  \mu \right)^2 + \Delta ^2} \right)  =0 .
\label{gap1d}
\een
Hence, since $\Delta $ is the order parameter of the difermion condensation phase transition, the solution of the gap equation (\ref{gap1d}) is useful in the characterization of the $(T,\mu, L)$-dependent phase diagram of the difermion sector of the model.

\begin{figure}[th]
\includegraphics[{height=8.0cm,width=8.0cm}]{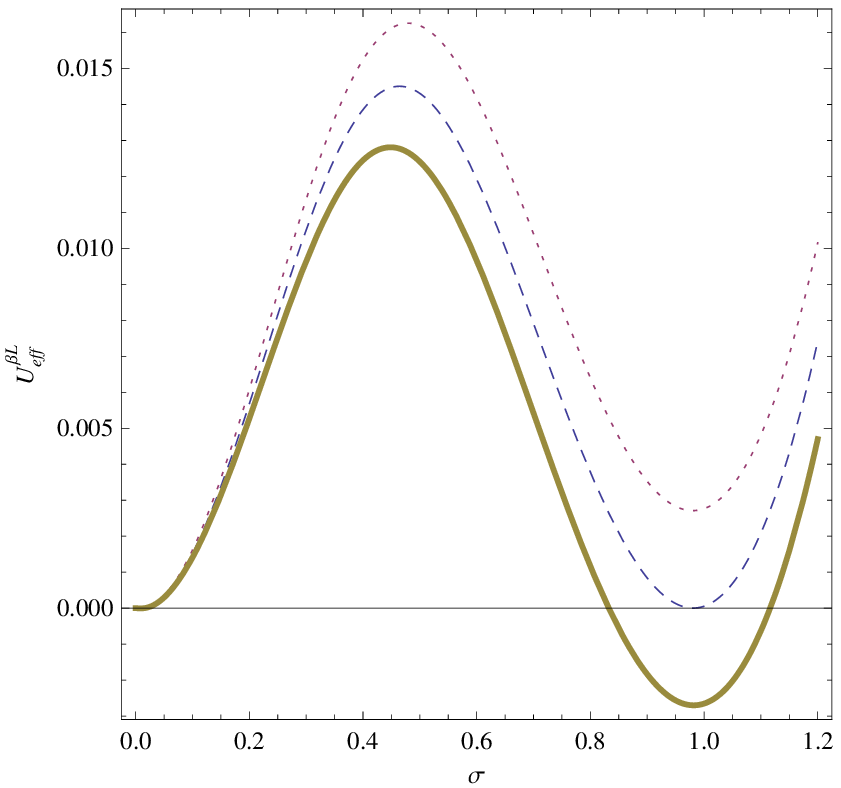}
\caption{Effective potential for the chiral condensate sector 
for  $x = 1/L = 0.2$ and $T=0.1$. Solid, Dashed and dotted
lines represent $\mu = 0.632, 0.639$ and $0.645$, respectively. } \label{fig1}
\end{figure}

\begin{figure}[th]
\includegraphics[{height=8.0cm,width=8.0cm}]{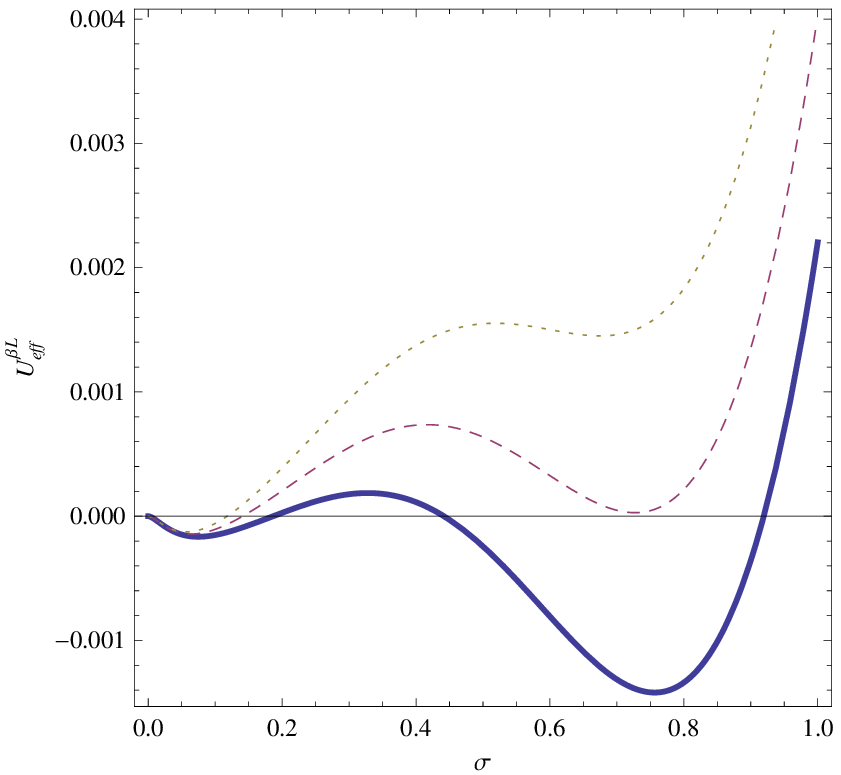}
\caption{Effective potential for the chiral condensate sector 
for  $x = 1/L = 0.3$ and $T=0.2$. Solid, dashed and dotted
lines represent $\mu = 0.63, 0.64$ and $0.65$, respectively. } \label{fig2}
\end{figure}

\begin{figure}[th]
\includegraphics[{height=8.0cm,width=8.0cm}]{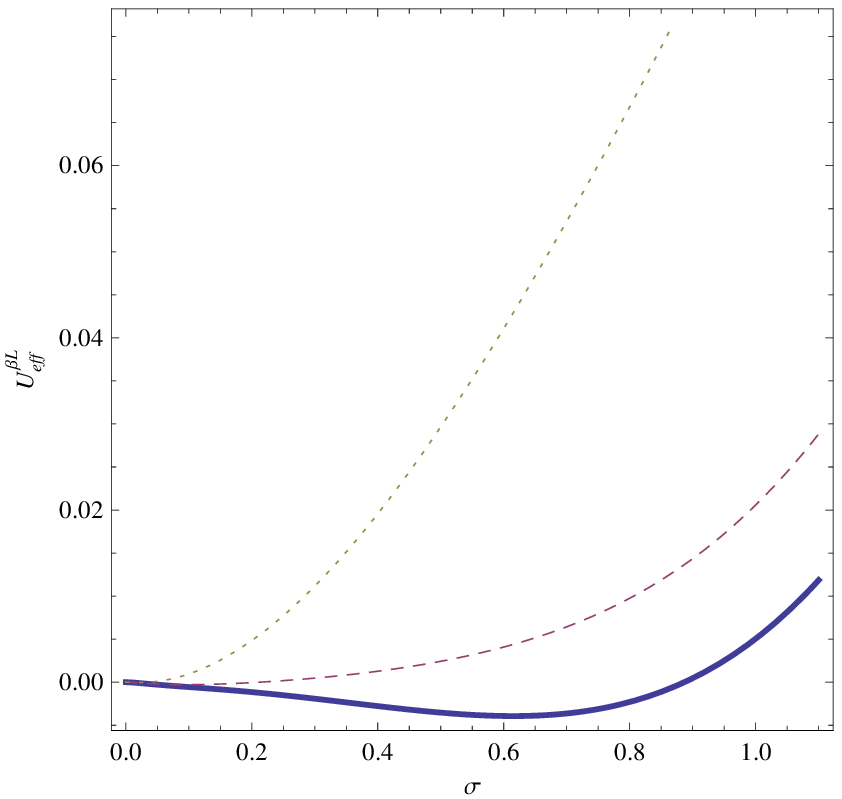}
\caption{Effective potential for the chiral condensate sector 
for  $x = 1/L = 0.4$ and $T=0.4$. Solid, dashed and dotted
lines represent $\mu = 0, 0.4$ and $0.8$, respectively. } \label{fig3}
\end{figure}

\begin{figure}[th]
\includegraphics[{height=8.0cm,width=8.0cm}]{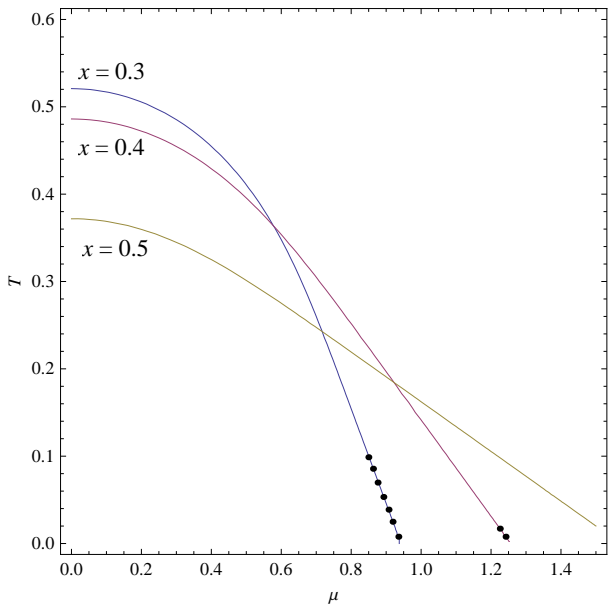}
\caption{The phase diagram for the chiral phase transition in the $(\mu,T)$-plane. 
As indicated, the curves are drawn for $x = 1/L = 0.3, 0.4$ and $0.5$, respectively. The chiral condensate region is below each curve. Dotted parts of the curves correspond to a first-order phase transition, while the solid ones represent a second-order transition. } \label{fig4}
\end{figure}

\section{Phase structure}


Now we are able to analyze the $L$-dependent critical curves in the
phase diagram of the model. First, we analyze the effective potential for the chiral condensate sector. Remarking that $\sigma _0 $ is the scale of
the theory, we can redefine the relevant quantities as $\Delta _0 / \sigma _0 \rightarrow \Delta _0 $, $L \sigma
_0\rightarrow L$, $ \mu / \sigma _0 \rightarrow \mu$ and $\beta
\sigma _0 \rightarrow \beta$.

In Figs. \ref{fig1}-\ref{fig3} are plotted the effective potential in Eq. (\ref{pot4}) for different values of $\mu, x = 1/L$ and $T$. 
Some detailed information on the influence of spatial boundaries can be obtained  from these figures; from Fig.\ref{fig1} we see that a first order transition occurs for increasing values of $\mu$, for the values of $L$ and $T$ given (in arbitrary units). For a smaller value of $L$ and a larger value of $T$, (see Fig.~\ref{fig2}), the phase transition takes place in two steps, as $\mu $ is increased: it is a  first order transition, but such that the absolute minimum of the effective potential is displaced to a smaller value of $\sigma$, as $\mu $ is increased ($\mu=0.63$ and $\mu=0.64$ in the figure); for a high enough value of $\mu$ ($\mu\geq 0.65$ in the figure), the transition becomes a second-order one, when the first non-vanishing extremum disappears \cite{IKM}. Finally, we see from Fig.~\ref{fig3} that there is a second order phase transition as $\mu$ increases,  for large enough values of $L$ and $T$.

To obtain the phase diagram more precisely in the case of a second-order transition, we must determine the values of $L, T$ and $ \mu$ at which the dynamical fermion mass vanishes. These critical lines are determined by taking the fermion mass equal to $zero$ in the gap equation,
\ben
\left. \frac{\partial}{\partial \sigma}  \bar{U}_{eff} ^{\beta,L} (\sigma) \right|_{\sigma = 0} = 0; 
\een
on the other hand, in the case of a first-order transition, we have studied the non-vanishing solution of the gap equation (\ref{gap1s}), togheter with the behavior of the effective potential. 

Taking into account simultaneously both first and second-order transitions, the critical lines in the $(T,\mu)$ plane are displayed in Fig.~\ref{fig4} for different values of $x=1/L$. In this figure, solid lines correspond to second-order and dotted lines correspond to first-order phase transitions, respectively. We see that as $x= 1/L$ increases, smaller values of temperature are necessary to reach the symmetric chiral phase in the region of lower values of $\mu$; in addition, in the region of greater values of $\mu$ and small $T $, the chiral broken phase spreads out. In what concerns  the order of the phase transition, we see that as $x= 1/L$ is increased, the region of first-order transition is suppressed, occurring only a second-order one.


\begin{figure}[th]
\includegraphics[{height=8.0cm,width=8.0cm}]{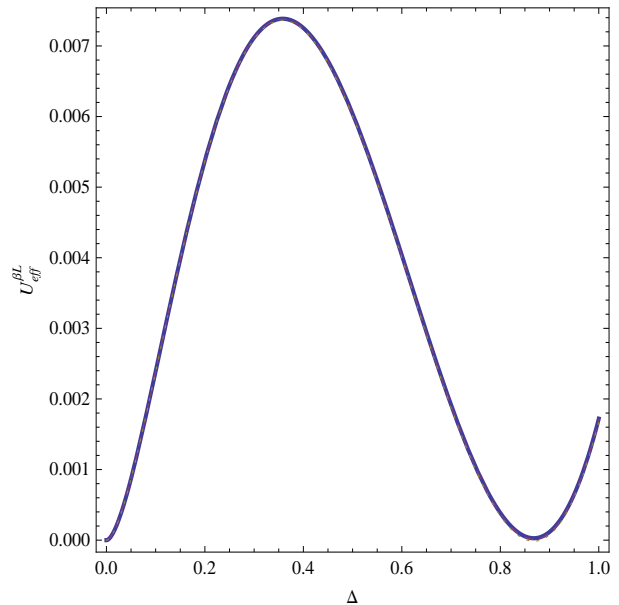}
\caption{Effective potential for the difermion condensate sector. 
for  $x = 1/L = 0.1$ and $T=0.294$. The lines representing $ \mu = 0, 0.5$ and $1$ coincide. } \label{fig5}
\end{figure}

\begin{figure}[th]
\includegraphics[{height=8.0cm,width=8.0cm}]{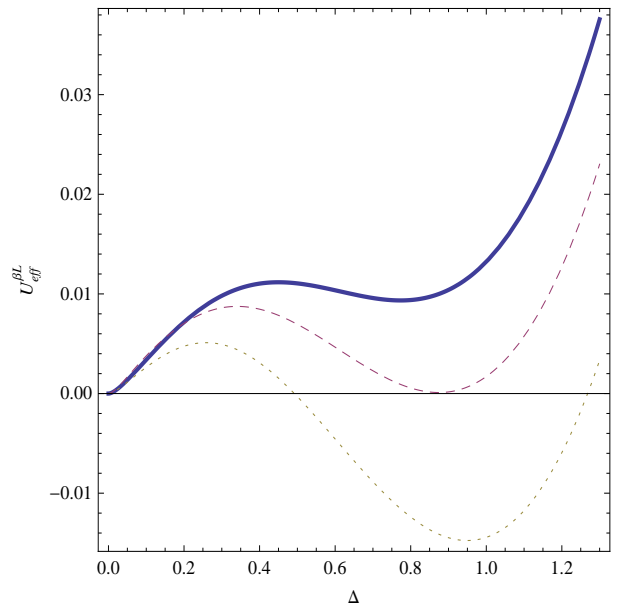}
\caption{Effective potential in the difermion condensate sector  
for  $x = 1/L = 0.5$ and $T=0.2$. Solid, dashed and dotted
lines represent $ \mu = 0.700, 0.747$ and $0.800$, respectively. } \label{fig6}
\end{figure}

\begin{figure}[th]
\includegraphics[{height=8.0cm,width=8.0cm}]{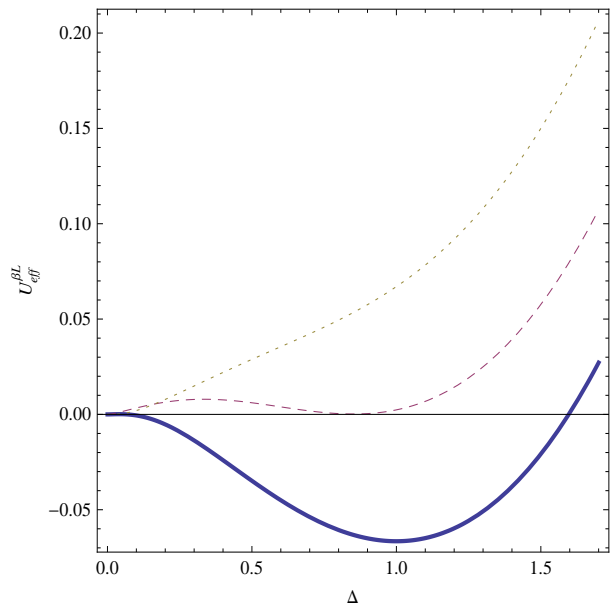}
\caption{Effective potential for the difermion condensate sector 
in the antiperiodic case 
Solid, dashed and dotted
lines represent $x = 1/L = 0.1, 0.49$ and $0.6$, respectively. } \label{fig7}
\end{figure}

\begin{figure}[th]
\includegraphics[{height=8.0cm,width=8.0cm}]{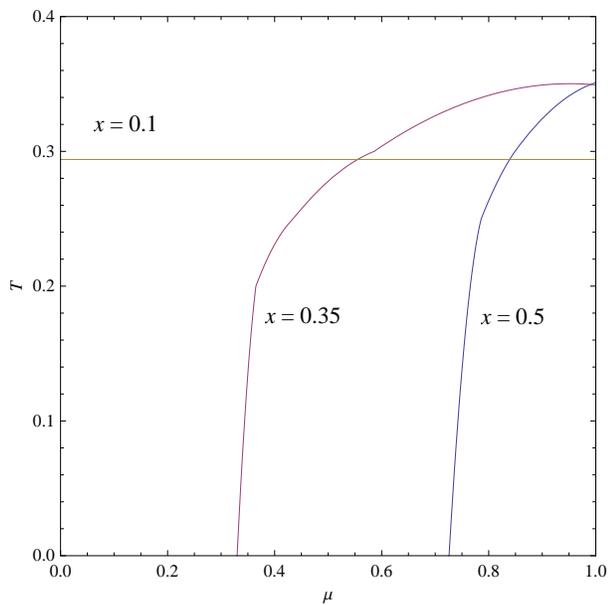}
\caption{The phase diagram for the difermion condensation phase transition in the $(\mu,T)$-plane. 
The lines represent $x = 1/L = 0.1, 0.35$ and $0.5$, respectively. The difermion condensate region is below each line, which represents a first-order phase transition.
} \label{fig8}
\end{figure}


Now let us analyze the fermion-fermion pairing phase structure. We plot in Figs. \ref{fig5}-\ref{fig7} the effective potential in Eq. (\ref{pot6}) for different values of $\mu, x = 1/L$ and $T$. 
From Fig.~\ref{fig5} we see that for small values of $x=1/L$,  that this phase transition is of first-order  and is independent of $\mu$ in these cases.  Fig.~\ref{fig6} shows that for higher value of $x$ and smaller temperature, the system undergoes a first order phase transition for increasing values of $\mu$.  From Fig.~\ref{fig7},  it can be seen that for fixed $\mu$ and a fixed smaller temperature, a first-order phase transition occurs as  $\mu$ increases. In these cases the difermion condensate phase is destroyed by experencing  a first-order phase transition as the size decreases or the chemical potential increases.


In Fig.~\ref{fig8} is plotted the phase diagram of the system in the
$(T,\mu)$-plane with different values for $x = 1/L$. From the figure we see that as the size of the system decreases, greater values of the chemical potential are necessary to
reach the difermion condensed phase region. On the other hand, for
larger sizes (for instance $L=(0.1)^{-1}$ in the figure), this region spreads out, and can be reached for practically any value of the chemical potential.

\section{Concluding Remarks}

In this article we have analysed finite-size effects for a version of the NJL model presented originally in Ref.~\cite{CMC} for finite temperature and density in free space. This has been done by using the extended Matsubara prescription and $zeta$-function regularization methods. Finite chemical potential effects have also been included. These techniques allowed us to construct a renormalized effective potential taking into account simultaneously the dependence on temperature, the size of the system and chemical potential. From this effective potential we have obtained the size-dependent gap equations at finite temperature and density. A throughout analysis of the size dependence of the phase diagram has been performed for both chiral and difermion sectors.

In what the chiral condensate sector is concerned, a first-order transition occurs for increasing values of the chemical potential at fixed values of temperature and of the size of the system. For smaller values of $L$ and simultaneously larger temperatures, the system undergoes a phase transition in two steps, for increasing values of the chemical potential: a first-order transition but such that the absolute minimum is displaced to a smaller value of $\sigma$, as $\mu$ increases. For some larger value of $\mu$ the transition becomes a second-order one. We have taken into account simultaneously first- and second-order transitions. This has been done by a study in the $T-\mu$ plane of the behavior of the system for different values of $x=1/L$. We conclude that as $x$ increases the region of a first-order transition disappears, leaving place for a second-order transition. 

The fermion-fermion pairing phase structure has also been investigated, from an analysis of the effective potential for different values of $\mu$, $x$ and $T$. We conclude that for small values of $x$ we have a first-order transition, independently of the value of $\mu$. For fixed $\mu$ and a smaller temperature, a first-order transition occurs as $x$ increases. In general, as the size of the system becomes smaller, larger values of $\mu$ are needed to attain the difermion condensate sector.

It should also be noticed that Fig~\ref{fig8} indicates that as $L$ decreases there exists a minimal size of the system for which the difermion condensate exists at zero chemical potential. For smaller values of $L$, the difermion phase reappears if we increase the chemical potential. The full phase structure, considering both sectors simultaneously, could note be obtained analytically.

{\bf Acknowledgments}

This work received partial financial support from CNPq (Brazil). A.P.C.M. thanks FAPERJ (Brazil) for partial finantial support.

\end{document}